\documentclass[sigconf,nonacm]{acmart}

\AtBeginDocument{%
  }

\usepackage{xcolor}
\usepackage{listings}
\usepackage[noabbrev]{cleveref}
\usepackage{epigraph}

\definecolor{darkslateblue}{RGB}{72,58,148}
\definecolor{forestgreen}{RGB}{57,127,50}
\definecolor{maroon}{RGB}{104,23,0}

\lstnewenvironment{code}
  {
  }
  {
  }

\lstnewenvironment{notcode}
  {
  }
  {
  }
  
\newcommand{\cd}[1]{\lstinline{#1}}

\begin{document}

\title{The Foil: Capture-Avoiding Substitution With No Sharp Edges}

\author{Dougal Maclaurin}
\email{dougalm@google.com}
\affiliation{%
  \institution{Google Research}
  \country{}
}
\author{Alexey Radul}
\email{axch@google.com}
\affiliation{%
  \institution{Google Research}
  \country{}
}
\author{Adam Paszke}
\email{apaszke@google.com}
\affiliation{%
  \institution{Google Research}
  \country{}
}

\begin{abstract}
Correctly manipulating program terms in a compiler is surprisingly difficult because of the need to avoid name capture.
The rapier from \citet{secrets} is a cutting-edge technique for fast, stateless capture-avoiding substitution for expressions represented with explicit names.
It is, however, a sharp tool---its invariants are tricky and need to be maintained throughout the whole compiler that uses it.
We describe the foil, an elaboration of the rapier that uses Haskell's type system to enforce the rapier's invariants statically, preventing a class of hard-to-find bugs, but without adding any run-time overheads.
\end{abstract}

\maketitle

\epigraph{There are only two hard things in Computer Science: cache invalidation and naming things.}{Phil Karlton}

\section{Introduction}

Names turn out to be one of the Hard Things in writing compilers as well.
In the Dex compiler, for instance, we've been following GHC's version of the Barendregt convention, ``the rapier'' \citep{secrets}.
It‘s elegant and it’s fast. It's also stateless, which is crucial for caching and concurrency.

But it‘s \emph{really} easy to screw up.
We’ve messed it up again\footnote{\scriptsize\url{https://github.com/google-research/dex-lang/commit/e979cae84c9b0cd612bed1013cdecf71e8c0d917}}
and again\footnote{\scriptsize\url{https://github.com/google-research/dex-lang/commit/a6425c60a70b5db8871ea05185c30995d24bfbb8}}
and again\footnote{\scriptsize\url{https://github.com/google-research/dex-lang/commit/c34ff0865306198aa9ed0c9ae1949325b6754dd7}}
and again\footnote{\scriptsize\url{https://github.com/google-research/dex-lang/commit/b96dbddba09bbd1e84f988da597bb350892c7fbd}}
and again\footnote{\scriptsize\url{https://github.com/google-research/dex-lang/commit/82c7edbde29c43e66eda8657de8853752709f11b}}
and again.\footnote{\scriptsize\url{https://github.com/google-research/dex-lang/commit/c154995fa5eea42acef69d39b8247da5e455c8c1}}
This had become one of the biggest barriers to implementing new language ideas and onboarding new people.
Worse, it made us hesitate to use name-based indirection in places it would have been helpful.

Here, we describe a design, ``the foil'', that implements the same naming discipline as the Simons' rapier but enforces it using Haskell's type system.
The design adds a phantom type parameter to every AST, representing the set of allowable free variables.
We keep the benefits of the rapier---speed and statelessness---but we make it harder to poke your eye out.

Anecdotally, after adopting the foil, formerly pervasive name handling bugs disappeared from the Dex compiler.
We have significantly more confidence in the quality of our implementation.
Much more of our time is spent on a productive discussion with the compiler, instead of puzzle-solving through ad-hoc debugging.

\section{Reviewing the Rapier}

The rapier \citep{secrets} is a discipline for fast and stateless name management.
The rapier applies when we are working with explicit names, as opposed to De Bruijn indices \citep{de-bruijn} or another expression representation such as locally-nameless \citep{locally-nameless} or higher-order abstract syntax~\citep{hoas}.

The canonical example task for name management is capture-avoiding substitution.
To implement this in rapier style we maintain, in addition to the substitution itself, the \emph{scope}, which is the set of variables that might appear free in the result.
In particular, all free variables of the terms we are substituting with should be contained in that scope.

The scope serves two purposes.
When performing substitution under a binder, the bound name might need to get refreshed so as to avoid capturing the free variables of terms in the substitution.
Having a scope lets us easily check whether a name might occur in those free variables, and with a well-chosen representation lets us efficiently generate fresh names if needed.
We therefore need neither a global name supply, nor to traverse terms repeatedly to compute their free variables, since the scope can be cheaply maintained during the substitution traversal.

Furthermore, not only can we refresh the binder, we can also be lazy about it!
Having access to the scope, we can deduce that certain binders are guaranteed to not capture any relevant variables, and we can avoid renaming them at all.

Using an explicit scope, capture-avoiding substitution on a simple expression language looks like \Cref{fig:rapier}, with straightforward backing data structures such as in \Cref{fig:raw_names}.

\begin{figure}
\begin{notcode}
substitute :: Scope -> Substitution Expr -> Expr -> Expr
substitute scope subst = \case
  Var v -> case lookup subst v of
    -- missing names imply identity substitution
    Nothing -> Var v
    Just x  -> x
  App f x -> App (recur f) (recur x)
    where recur = substitute scope subst
  Lam v body -> if occursIn v scope
    then
      let v' = freshWrt scope
          subst' = extendSubst v (Var v') subst
          scope' = extend v' scope in
      Lam v' (substitute scope' subst' body)
    else
      Lam v  (substitute (extend v scope) subst body)
\end{notcode}
\caption{Substitution wielding the rapier}{on a simple expression language; but already shows multiple sharp edges.}
\label{fig:rapier}
\end{figure}

The advantages of the rapier are substantial:
\begin{itemize}
    \item It's fast: a multi-name capture-avoiding substitution happens in a single pass over the input \cd{expr}, and we often do not even need to traverse the terms being substituted for their free variables, because we may already have the scope on hand when we start.
    \item It's stateless (no name supply), so it's parallelizable \linebreak[4] and cacheable.
    \item We do not change names that are already fresh, so substituting with the empty substitution does not change the term.
\end{itemize}

Unfortunately, it's also very easy to get wrong.
Here are four obvious sharp edges:

\begin{itemize}
    \item In the \cd{Var} case, we could just return \cd{Var v} instead of first checking to see if it's in the substitution.
    \item In the \cd{App} case, we could forget to apply the substitution to \cd{f} or \cd{x}, or we could apply it more than once. We could also make a similar error in the \cd{Lam} case.
    \item In the \cd{Lam} case, we could forget to extend the scope or extend it incorrectly.
    \item In the \cd{Lam} case, we could forget to extend the substitution or extend it incorrectly.
\end{itemize}

Substitution itself is the simplest illustration of the problem, but the real headaches come from the more complicated passes: type inference, normalization, linearization (for automatic differentiation), transposition (also for automatic differentiation), optimizations, lowering to imperative IRs, and so forth.
These all include logic that looks a lot like substitution. For example, the linearization pass carries an environment that maps each name in the input program to the corresponding primal and tangent terms for the output program.
Alongside the actual logic for each pass we have to deal with the petty bureaucracy of name scopes and renaming to avoid clashes. That's where we make mistakes.

And the situation is far worse for the abstract syntax of a large language like Dex, which has dozens of constructors instead of just three.
The trickiest are those that introduce new binders and lexical scopes: lambda, pi types, let bindings, ``for'' expressions, case expressions, effect handlers, AD transformations, and dependent data constructors.
Some have nested lists of scoped binders, like the binders in a dependent data constructor pattern.
%

Dex is also dependently typed, such that the names that can appear in types are in the same namespace as the names that appear in terms.
So even a type-preserving compiler pass can nonetheless trigger renames in types (to avoid name capture), and other such examples; so one has to pay attention to applying substitutions in places that have nothing to do with the business logic of the code transformation one is trying to implement.  All while taking care not to apply each substitution more than once.

\begin{figure}
\begin{code}
import qualified Data.IntSet as Set
import qualified Data.IntMap as IM

-- The Int is an ID, not a De Bruijn index
newtype RawName = RawName Int
  deriving (Eq, Ord)
newtype RawScope = RawScope Set.IntSet
  deriving (Eq)

rawEmptyScope :: RawScope
rawEmptyScope = RawScope Set.empty

rawFreshName :: RawScope -> RawName
rawFreshName (RawScope s) | Set.null s = RawName 0
                          | otherwise = RawName (Set.findMax s + 1)

rawExtendScope :: RawName -> RawScope -> RawScope
rawExtendScope (RawName i) (RawScope s) = RawScope (Set.insert i s)

rawMember :: RawName -> RawScope -> Bool
rawMember (RawName i) (RawScope s) = Set.member i s

newtype RawSubst a = RawSubst (IM.IntMap a)

rawIdSubst :: RawSubst a
rawIdSubst = RawSubst IM.empty

rawLookup :: RawSubst a -> RawName -> Maybe a
rawLookup (RawSubst env) (RawName i) = IM.lookup i env

rawExtendSubst :: RawName -> a -> RawSubst a -> RawSubst a
rawExtendSubst (RawName i) val (RawSubst env) =
  RawSubst (IM.insert i val env)
\end{code}
\caption{Raw names, scopes and substitutions}{
that the foil makes safer to use.
Here we \cd{newtype}-wrap a representation of names as uninterpreted \cd{Int}s, but anything that makes a good map key will work.
}
\label{fig:raw_names}
\end{figure}

What‘s more, bad substitutions are the worst kinds of errors. If we’re lucky, the buggy compiler pass will produce an ill-typed object program that we’ll catch at the next internal type checking step. More often, we just get very confusing behavior in a downstream pass. It's also hard to find minimal reproducers and small tests for these sorts of errors, because whether you get a name collision depends on all the other names in the program.

So, what would a safer rapier look like?

\section{Forging the Foil}

The big idea of the foil is to annotate the types of expressions with a type parameter indicating the in-scope variables that may occur in the expression, \cd{Expr n}.
For a given \cd{n}, there is only one \cd{Scope n}, the actual in-scope set.

The type system will guarantee that if \cd{x::Scope n} and \cd{y::Scope n} then \cd{x == y}, and also that any variables occurring in an expression \cd{Expr n} appear in that \cd{Scope n}.
With that guarantee, if we find a name that doesn't occur in \cd{Scope n}, we know it doesn't occur in \cd{Expr n}, without needing to check.

Likewise, we will define structures like substitutions to ensure that we can't forget to extend them appropriately.

\subsection{Safer Scopes}
\label{sec:safer-scopes}

We start by defining
\begin{enumerate}
\item a kind \cd{S} (for ``Scope'') of scope-indexing types,
\item an \cd{S}-indexed type \cd{Name n}, representing a name subject to the foil, and
\item another \cd{S}-indexed type \cd{Scope n}, representing a scope containing exactly the inhabitants of type \cd{Name n}.
\end{enumerate}

By only allowing \cd{Name n} and \cd{Scope n} objects to be created in a limited set of ways, we enforce the Scope Invariant:
\begin{definition}[Scope Invariant]
For every \cd{n} of kind \cd{S}:
\begin{itemize}
    \item If \cd{x :: Scope n} and \cd{y :: Scope n} then \cd{x == y}.
    \item If \cd{x :: Scope n} and \cd{name :: Name n} then \cd{member name x} is true.
\end{itemize}
\end{definition}

We will mostly be introducing \cd{S}-kinded variables with rank-2 polymorphism, but we do use one constructor for the \cd{S} kind to get started.
A \cd{Scope VoidS} is the empty scope that contains no names, and the type \cd{Name VoidS} is uninhabited.

\begin{notcode}
data S = VoidS
\end{notcode}

\begin{figure}
\begin{code}
data S = VoidS

newtype Name (n::S) = UnsafeName RawName
  deriving (Eq, Ord)

newtype Scope (n::S) = UnsafeScope RawScope
  deriving (Eq)

emptyScope :: Scope VoidS
emptyScope = UnsafeScope rawEmptyScope

member :: Name l -> Scope n -> Bool
member (UnsafeName name) (UnsafeScope s) = rawMember name s
\end{code}
\caption{Safer names and scopes}{and basic operations on them}
\label{fig:safer-names-1}
\end{figure}

While our type indexing statically proves that every \cd{Name n} is a member of \cd{Scope n}, we do also need a runtime representation for names and scopes---for instance, a \cd{Name l} may shadow a name in \cd{Scope n}, and we need to be able to check whether that happens.

We reuse the raw name and scope operations from \Cref{fig:raw_names}.
For simplicity, we implement \cd{Name}s as \cd{Int}s at runtime here, but of course a more sophisticated system could store any desired information in them.
For practical use, we strongly recommend extending the APIs we present with a notion of ``name hints'', which are the information that one would like a name to carry other than its identity.%
\footnote{In the Dex compiler, runtime names are integerss, but we pack the first several characters of the user-facing variable name into the high bits of the integer to make intermediate representation dumps more readable.  This also reduces renaming churn, because user-facing variable names are usually already locally unique.}

So, how do we type the raw operations from \Cref{fig:raw_names} to maintain the scoping invariant?
The first step, in \Cref{fig:safer-names-1}, is to add the phantom \cd{S}-kinded type parameter to the raw representation.
The empty scope gets tagged \cd{VoidS}, proving to the type system that it is, indeed, empty, and membership testing can ignore the phantoms because it's always safe.

The interesting operation is creating a fresh name.
If Haskell supported existential types, we could type it as
\begin{notcode}
-- Need existentials
freshName :: Scope n -> (exists l. Name l)
\end{notcode}
but as it stands, we have to transform it to continuation-passing style
\begin{code}
withFreshCPS :: Scope n -> (forall l. Name l -> r) -> r
\end{code}

That's still not good enough, though, because we also want to be able to create a \cd{Scope l} that includes the new \cd{Name}, while proving that it cannot include any other names.
For this, \Cref{fig:name-binders} introduces another new type
with its own invariant:
\begin{definition}[Binder Invariant]
A \cd{NameBinder n l} only exists if the scope indexed by \cd{l} extends the scope indexed by \cd{n} by exactly the single name contained in the \cd{NameBinder n l}.
\end{definition}

\begin{figure}
\begin{code}
-- n is the scope above the binder
-- l (for "local") is the scope under the binder
newtype NameBinder (n::S) (l::S) = UnsafeBinder (Name l)

nameOf :: NameBinder n l -> Name l
nameOf (UnsafeBinder name) = name

extendScope :: NameBinder n l -> Scope n -> Scope l
extendScope (UnsafeBinder (UnsafeName rn)) (UnsafeScope s) =
  UnsafeScope (rawExtendScope rn s)

withFreshBinder :: Scope n
  -> (forall l. NameBinder n l -> r) -> r
withFreshBinder (UnsafeScope rs) cont = cont binder where
  binder = UnsafeBinder (UnsafeName (rawFreshName rs))
\end{code}
\caption{Name binders}{and how to safely allocate names and extend scopes.
Here and throughout, the annotated types of unsafely-implemented foil operations are more restrictive than would be inferred---indeed, the type annotations are what make them safe to use!
}
\label{fig:name-binders}
\end{figure}

The binder invariant is what allows a \cd{Scope n} to be extended safely: a \cd{NameBinder n l} object can only be created by \cd{withFreshBinder} at index \cd{n}, so it carries the proof that the new \cd{Scope l} is unique and satisfies the Scope Invariant.

\subsection{Safer Expressions}

Now that we have names and scopes obeying the Scope Invariant, we can use them to define well-scoped expressions.
A well-scoped expression is one that obeys the Expression Invariant:
\begin{definition}[Expression Invariant]
For \cd{n} of kind \cd{S}, every expression of type \cd{Expr n} has free variables contained in \cd{Scope n} (which, as we recall, is uniquely determined by the index \cd{n}).
\end{definition}

The expression type is of course specific to the object language to be implemented with the foil; and this is actually a place where the user has to take care to define their \cd{Expr} ADT to actually obey that invariant.
The benefit from the foil, though, is that the invariant follows (or doesn't follow) from the data definition for \cd{Expr}, and uses of it will be scope-correct as long as they type-check in Haskell.

What does a user have to do to make sure their expression type enforces the Expression Invariant?
Just three rules:
\begin{enumerate}
    \item Free variables are exposed as \cd{Name}s parameterized with their scope;
    \item Binders are exposed as \cd{NameBinder}s; and
    \item Sub-expressions (that might have free variables) are parameterized by their scope.
\end{enumerate}

For example, a simple untyped lambda calculus has all those constructs and pretty much nothing else:
\begin{code}
data Expr n =
    Var (Name n)
  | Lam (LamExpr n)
  | App (Expr n) (Expr n)

data LamExpr n where
  LamExpr :: NameBinder n l -> Expr l -> LamExpr n
\end{code}

Note that the binder form uses a GADT to existentially hide the local scope parameter \cd{l}.

Of course, if all you wanted to implement was the lambda calculus, you wouldn't need the foil.
\Cref{fig:example-dependent} shows an expression type for a more involved language.
\begin{figure}
\begin{code}
data Type n = TyVar (Name n)
  | TyInt | TyReal | TyFun (TyFunType n)

data TyFunType n where
  TyFunType ::
    { tyArg :: NameBinder n l
    , tyArgTy :: Type n
    , tyRetTy :: Type l
    } -> TyFunType n

data DepExpr n = DepVar (Name n)
  | IntLit Int | RealLit Float
  | DepLam (DepLamExpr n) | DepApp (Expr n) (Expr n)

data DepLamExpr n where
  DepLamExpr ::
    { arg :: NameBinder n l
    , argTy :: Type n
    , body :: DepExpr l
    , retTy :: Type l
    } -> DepLamExpr n
\end{code}
\caption{An example scope-indexed abstract syntax}{following the Expression Invariant called for by the foil.}
\label{fig:example-dependent}
\end{figure}
We see from the definition of \cd{Expr} that this is a dependently typed language, where the  type of a function's result is allowed to depend on the argument.
If we had written \cd{retTy :: Type n}, however, we would disallow this, because Haskell would prove that the name bound by \cd{arg} never appears in \cd{retTy}.

Likewise, if we wanted to statically enforce that, say, type variables and term variables occupied different name spaces, we could define a language where \cd{Expr} was parameterized by two \cd{S}-kinded variables, one for the term scope and one for the type scope.  Et cetera.

\subsection{Safer Sinking}
\label{sec:safer-sinking}

Now that we have established the Scope, Binder, and Expression Invariants that describe expressions at rest, let's consider expressions in motion.
To wit, suppose we are looking to substitute some term \cd{term :: Expr n} into some expression \cd{expr :: Expr n}.
If \cd{expr} is a non-binding form such as a \cd{Var} or an \cd{App} everything is fine, but what do we do when \cd{expr} is a binding form?
We now have in our hands a new scope parameter \cd{l}, a \cd{binder :: NameBinder n l}, and an inner expression \cd{Expr l}.
We can't just recur, because the new binder might capture a name from \cd{term}; and indeed, the foil prevents this potential mistake, because there is no way to insert \cd{term :: Expr n} into \cd{expr' :: Expr l}.

What we want is to check that the new name introduced by \cd{binder :: NameBinder n l} does not capture any of the free variables of \cd{term :: Expr n}, and then reinterpret it as \cd{term :: Expr l}.
We call such reinterpretation \emph{sinking}.
It merits its own discussion because it occurs all the time in compilers---whenever you want to interpret anything from higher in the expression tree in some lower context that may have more binders in scope, you have to sink it.

So, when is it safe to sink a \cd{term :: Expr n} to \cd{term' :: Expr l}?  We need
\begin{enumerate}
    \item Every name that appears free in \cd{term} must appear in \cd{Scope l}
    \item Every name that appears free in \cd{term} must mean the same thing in \cd{Scope l} that it does in \cd{Scope n}; in other words, names new to \cd{l} must not shadow bindings of names of \cd{term}.
\end{enumerate}
The free variables of \cd{term} are a runtime property, but we can over-approximate it statically by turning it into a property of scopes: if \cd{Scope l} contains \emph{all} the names of \cd{Scope n}, and the extension from \cd{n} to \cd{l} shadows \emph{none} of them, then \emph{any} \cd{term :: Expr n} is safe to sink to \cd{l}, regardless of its free variables.
This is the critical insight used in the rapier \citep{secrets}.

We represent these two properties separately as Haskell typeclasses.
First, we define a class \cd{Ext n l} that guards the Extension Invariant:
\begin{definition}[Extension Invariant]
If \cd{n :: S} and \cd{l :: S} have an instance of \cd{Ext n l}, then \cd{Scope n} is a subset of \cd{Scope l} (not necessarily strict, and not necessarily without shadowing).
\end{definition}

The \cd{Ext} class itself has no methods; we will just be using it to justify coercions in unsafe implementations of functions of the foil.
We do, however, take the opportunity to use its definition to have GHC automatically deduce reflexivity and transitivity of \cd{Ext}.
\begin{code}
class ExtEndo (n::S)

class (ExtEndo n => ExtEndo l) => Ext (n::S) (l::S)
instance (ExtEndo n => ExtEndo l) => Ext n l
\end{code}
The idea is that \cd{ExtEndo n => ExtEndo n} is always true, so GHC can synthesize \cd{Ext n n} for any \cd{n} on its own; and GHC can also synthesize \cd{Ext n1 n3} from \cd{Ext n1 n2} and \cd{Ext n2 n3} by hypothesizing \cd{ExtEndo n1}, deducing \cd{ExtEndo n3} from it, and concluding the implication.
This definitional trick isn't strictly necessary, but makes \cd{Ext} considerably more ergonomic in practice.

A user could break the foil by defining their own instances of \cd{Ext} (or \cd{ExtEndo}); please don't.
On the plus side, new instances are not needed to use the foil correctly, so there is little risk of accidental mis-use.

Second, we define a class \cd{Distinct n} that guards the Distinctness Invariant:
\begin{definition}[Distinctness Invariant]
If \cd{n :: S} has an instance of \cd{Distinct n}, then all the names in \cd{n} are distinct, i.e., none of them shadow any others.
\end{definition}

Like \cd{Ext}, the \cd{Distinct} class is also a method-less marker that the user need not (and must not) define their own instances for.
The definition is straightforward:
\begin{code}
class Distinct (n::S)
instance Distinct VoidS
\end{code}

Since \cd{Distinct} and \cd{Ext} often (but not always) travel together, we also define
\cd{DExt} as a constraint alias that means both of them.
\begin{code}
type DExt n l = (Distinct l, Ext n l)
\end{code}

We now have the machinery we need to type and define \cd{sink}:
\begin{code}
concreteSink :: DExt n l => Expr n -> Expr l
concreteSink = unsafeCoerce
\end{code}
One of the major advantages of using a name-based representation of expressions in the first place (as opposed to De Bruijn indices) is that sinking is free at runtime; all this machinery is about teaching Haskell's type system to keep track of when it's safe.%
\footnote{The \cd{S}-kinded type parameters are always phantom, so this coercion is entirely safe. If we allowed them to have a \cd{phantom} role, we could use the safe coercions from \citet{safe-coercions}, but they have to be declared as \cd{nominal} to make it impossible to accidentally change namespace parameters through safe APIs. Hence, we fall back to a safe unsafe coercion.}

In particular, we do not define a variant of \cd{sink} that would dynamically check the safety of a given sinking.
While such a variant is certainly possible, the place to handle a name clash is not at the point of sinking, but at the point where one can rename the offending binder to avoid the clash.
We thus prefer to define fresh-name producers to statically prove distinctness (\Cref{sec:safer-scopes-again}) and sinking to statically require it.

\subsection{Safer Scopes Again}
\label{sec:safer-scopes-again}

Where do we get instances of \cd{Ext} and \cd{Distinct}?
We provide them at the same time as we create new scope indices with rank-2 polymorphism.
To wit, we know \cd{l} is an extension of \cd{n} when we constructed it by adding names to \cd{n}; and we know \cd{l} is all-distinct when the names are fresh and \cd{n} was all-distinct to begin with.
We capture both of these with the type of our main name introduction function:
\begin{notcode}
withFresh :: forall n r. Distinct n => Scope n
  -> (forall l. DExt n l => NameBinder n l -> r) -> r
\end{notcode}
Where \cd{withFreshBinder} from \Cref{sec:safer-scopes} gave us a binder that was fresh at runtime, \cd{withFresh} also gives us a static proof that this binder is fresh.

\begin{figure}
\begin{code}
withFresh :: Distinct n => Scope n
  -> (forall l. DExt n l => NameBinder n l -> r) -> r
withFresh scope cont = withFreshBinder scope \binder ->
  unsafeAssertFresh binder cont

unsafeAssertFresh :: forall n l n' l' r. NameBinder n l
  -> (DExt n' l' => NameBinder n' l' -> r) -> r
unsafeAssertFresh binder cont =
  case unsafeDistinct @l' of
    Distinct -> case unsafeExt @n' @l' of
      Ext -> cont (unsafeCoerce binder)

data DistinctEvidence (n::S) where
  Distinct :: Distinct n => DistinctEvidence n

unsafeDistinct :: DistinctEvidence n
unsafeDistinct = unsafeCoerce (Distinct :: DistinctEvidence VoidS)

data ExtEvidence (n::S) (l::S) where
  Ext :: Ext n l => ExtEvidence n l

unsafeExt :: ExtEvidence n l
unsafeExt = unsafeCoerce (Ext :: ExtEvidence VoidS VoidS)

withRefreshed :: Distinct o => Scope o -> Name i
  -> (forall o'. DExt o o' => NameBinder o o' -> r) -> r
withRefreshed scope name cont = if member name scope
  then withFresh scope cont
  else unsafeAssertFresh (UnsafeBinder name) cont
\end{code}
\caption{Fresh names, in full}{including synthesizing instances for the \cd{Distinct} and \cd{Ext} marker classes.
Synthesizing preserves their invariants because the semantics of \cd{withFresh} guarantees them, and is safe at runtime because the classes have no methods.
Also a churn-reducing optimization: \cd{withRefreshed} is a variant of \cd{withFresh} that checks whether the given name is already fresh, and just reuses it.}
\label{fig:final-with-fresh}
\end{figure}

Note that with this type we cannot generate a static freshness proof with respect to a scope we do not statically know to be all-distinct.
This is a cost of using our relatively imprecise Distinctness Invariant, but it's not actually a very serious cost in practice.
Guaranteeing \cd{Distinct n} is not hard in a top-down traversal of a closed term.

Implementing \cd{withFresh} just reuses \cd{withFreshBinder}, except we also need a bit of \cd{unsafeCoerce} trickery to synthesize the \cd{Ext} and \cd{Distinct} classes for the continuation.
We can do that safely because they are just markers without methods.
The code appears in \Cref{fig:final-with-fresh}.
We also add a \cd{withRefreshed} variant of \cd{withFresh} (which must also be implemented unsafely) which reuses the underlying name if it's already fresh,
reducing renaming churn.

\subsection{Generic safe sinking}

The type we gave to \cd{concreteSink} in \Cref{sec:safer-sinking} is specialized to a concrete \cd{Expr} type.
But that type belongs to a hypothetical compiler being written using the foil, whereas the concept of sinking belongs to the foil system itself.

What's the right generalization?
Not all types of kind \cd{S -> *} are safe to sink---\cd{Expr} is OK, but \cd{Scope} is not.
The reason sinking \cd{Expr} is safe is that it's covariant in the scope index: an \cd{expr :: Expr n} contains references to a bunch of names in the \cd{n} scope; if we can understand all names from a larger (and non-shadowing) scope, then we can still understand all names in \cd{expr}.

We capture this constraint with a typeclass:
\begin{code}
class Sinkable (e :: S -> *) where
  sinkabilityProof :: (Name n -> Name l) -> e n -> e l

instance Sinkable Name where
  sinkabilityProof rename = rename

sink :: (Sinkable e, DExt n l) => e n -> e l
sink = unsafeCoerce
\end{code}

The \cd{sink} function is implemented as a coercion, so it never actually calls \cd{sinkabilityProof}; but conceptually, the logic is
\begin{enumerate}
    \item The \cd{Ext n l} instance tells us that the identity function on raw names can be safely typed \cd{idRename :: Name n -> Name l}.
    \item The \cd{Distinct l} instance tells us that this function is meaning-preserving.
    \item The \cd{Sinkable e} instance tells us that \cd{e} is covariant in the scope index.
    \item So \cd{sink expr} conceptually remaps every \cd{Name n} in \cd{expr} to the corresponding \cd{Name l}, but because we know their runtime representations are the same, that takes zero runtime work.
\end{enumerate}

Since the foil never calls any \cd{sinkabilityProof} methods, the client can just leave the implementation thereof \cd{undefined} if they convince themselves that a given type should be \cd{Sinkable}.
However, writing such proofs out and type-checking them adds a layer of safety.
Since safety is what the foil is all about, we present a sinkability proof for our example \cd{Expr} type in \Cref{sec:expr-sinkability}.

\section{Safer Substitution}

We are now ready to see how the foil helps us write substitution with fewer bugs.
How do we model names in a substitution?
We have an expression-like thing of some scope-indexed type, \linebreak[4] \cd{expr :: (e :: S -> *)}.
It may contain names of type \cd{Name i} (\cd{i} for ``input''), and we want to replace some of them with other terms, also of type \cd{e}.
To make sure we apply the substitution exactly once, we index the replacement terms by a (potentially) different scope, let's say \cd{o} for ``output''.
So applying substitution has this type:
\begin{code}
substitute :: Distinct o
           => Scope o -> Substitution e i o
           -> e i -> e o
\end{code}
(We require the output scope to be distinct because we may need to use \cd{withFresh} to allocate names that are fresh with respect to it, to avoid variable capture.)

\begin{figure}
\begin{code}
data Substitution (e::S -> *) (i::S) (o::S) =
  UnsafeSubstitution (forall n. Name n -> e n) (IM.IntMap (e o))

lookupSubst :: Substitution e i o -> Name i -> e o
lookupSubst (UnsafeSubstitution f env) (UnsafeName (RawName name)) =
  case IM.lookup name env of
    Just e  -> e
    Nothing -> f (UnsafeName (RawName name))

idSubst :: (forall n. Name n -> e n) -> Substitution e i i
idSubst f = UnsafeSubstitution f IM.empty

addSubst :: Substitution e i o -> NameBinder i i' -> e o
  -> Substitution e i' o
addSubst (UnsafeSubstitution f env)
  (UnsafeBinder (UnsafeName (RawName name))) e =
    UnsafeSubstitution f (IM.insert name e env)

addRename :: Substitution e i o -> NameBinder i i' -> Name o
  -> Substitution e i' o
addRename s@(UnsafeSubstitution f env)
  b@(UnsafeBinder (UnsafeName name1))
  n@(UnsafeName name2)
  | name1 == name2 = UnsafeSubstitution f env
  | otherwise = addSubst s b (f n)

instance (Sinkable e) => Sinkable (Substitution e i) where
  sinkabilityProof rename (UnsafeSubstitution f env) =
    UnsafeSubstitution f (fmap (sinkabilityProof rename) env)
\end{code}
\caption{Safer substitutions}{including the performance optimization of eliding names mapped to themselves.
Now this optimization has to be under the \cd{Substitution} abstraction boundary, because implementing it requires unsafe manipulation of the scope indices.
Without this optimization the API can be simpler, for instance eliding \cd{addRename}, but we include it to make sure the foil can replicate the rapier's performance tricks.}
\label{fig:safer-substitutions}
\end{figure}

The \cd{Substitution} data structure needs to carry the actual map of input names to terms being substituted for them.
We also add a performance optimization for the common case where a name is being changed to itself.  This happens when the substitution occurs in a local scope (e.g., in the scope of the top-level environment of the object language), and when going under a binder that turned out to be non-shadowing at runtime (\cd{withRefreshed} from \Cref{sec:safer-scopes-again}).
In this case, we will not add it to the runtime substitution, and instead just coerce it to the output index.
But for that to suffice, we also need to be able to inject unchanged names into the type \cd{e o} of substitution results, for which purpose the \cd{UnsafeSubstitution} constructor accepts a \cd{Name n -> e n} function that can be applied at any index \cd{n}.
For the \cd{Expr} language we are using as our running example, that function would be \cd{Var}.

\Cref{fig:safer-substitutions} defines the \cd{Substitution} type and implements the basic operations on it, taking care to maintain the Substitution Invariant:
\begin{definition}[Substitution Invariant]
If a \cd{Substitution e i o} exists, then it gives semantics to every \cd{name :: Name i}, and \cd{lookupSubst} implements those semantics.
Namely,
\begin{enumerate}
    \item If \cd{name} was last added with an \cd{addSubst} whose third argument was \cd{e}, then \cd{lookupSubst} returns \cd{sink e}.
    \item If \cd{name} was last added with an \cd{addRename} whose third argument was \cd{name'}, then \cd{lookupSubst} returns \cd{f name'}.
    \item Otherwise, \cd{name} was present in the root scope at which we called \cd{idSubst} and \cd{lookupSubst} returns \cd{f (unsafeCoerce name)}.
\end{enumerate}
\end{definition}

With those (unsafely implemented) pieces, we can now code substitution on any expression type that obeys the Expression Invariant, and the foil will prevent us from making mistakes.
For example, \Cref{fig:foil} implements substitution on the \cd{Expr n} lambda calculus we've been working with.
\begin{figure}
\begin{code}
substituteExpr :: Distinct o => Scope o -> Substitution Expr i o
  -> Expr i -> Expr o
substituteExpr scope subst = \case
  Var name -> lookupSubst subst name
  App f x -> App (recur f) (recur x)
    where recur = substituteExpr scope subst
  Lam (LamExpr binder body) ->
    withRefreshed scope (nameOf binder) (\binder' ->
      let subst' = addRename (sink subst) binder (nameOf binder')
          scope' = extendScope binder' scope
          body'  = substituteExpr scope' subst' body in
      Lam (LamExpr binder' body'))
\end{code}
\caption{Substitution wielding the foil}{on the simple object language from \Cref{fig:rapier}, except that the foil calls for the \cd{Expr n} type to be well-indexed.
Otherwise the code is essentially the same, just harder to cut oneself on.}
\label{fig:foil}
\end{figure}

How does the typing discipline help us get this right?
\begin{itemize}
    \item We can't forget to recur in \cd{App} because the only way to make an \cd{Expr o} out of an \cd{Expr i} is to call \cd{substituteExpr scope subst}.
    \item We can't recur more than once on the same term because \cd{substituteExpr} with \cd{subst} doesn't accept \cd{Expr o} as input.
    \item We can't forget to look up a \cd{Var} because the only way to get an \cd{Expr o} out of a \cd{Name i} is \cd{lookupSubst}.  (Notably, if we wrote \cd{Var name} for that case, that would be an \cd{Expr i}.)
    \item When we go under a binder, the body doesn't have type \cd{Expr i} any more, but rather \cd{Expr i'} for some index \cd{i'} that was existentially hidden in the \cd{LamExpr}.
    So we can't forget to extend \cd{subst}, because we need a \cd{Substitution e i' something} to recur.
    \item The only way to get a \cd{Substitution e i' something} is with either \cd{addSubst} or \cd{addRename}.  We wouldn't use \cd{addSubst} because we have no plausible \cd{e o} to pass to it.  That leaves \cd{addRename}, which requires a \cd{NameBinder o o'}. The input binder has type \cd{NameBinder i i'}, so we can't forget to use \cd{withFresh} to create a new binder in the output scope.
    \item Extending the substitution with \cd{addRename} changes the type of the output scope to \cd{o'} as well, so we can't forget to extend the \cd{scope} argument (which would otherwise still have type \cd{Scope o}).
    \item We also can't accidentally rebuild the result \cd{Lam} with the input \cd{binder}, because the returned \cd{body'} has type \cd{Expr o'}, and the only binder that is directly above it (as required by the \cd{LamExpr} GADT) is \cd{binder'}.
\end{itemize}

For example, here's the error message that GHC version 8.10.7 emits when we modify the \cd{substituteExpr} function from \Cref{fig:foil} to try to erroneously use \cd{subst} instead of \cd{subst'}:
\begin{verbatim}
main.lhs:834:47: error:
- Couldn't match type 'l' with 'i'
  'l' is a rigid type variable bound by
    a pattern with constructor:
      LamExpr :: forall (n :: S) (l :: S).
            NameBinder n l -> Expr l -> LamExpr n,
    in a case alternative
    at main.lhs:830:8-26
  'i' is a rigid type variable bound by
    the type signature for:
      substituteExpr :: forall (o :: S) (i :: S).
            Distinct o => Scope o
                    Substitution Expr i o -> Expr i -> Expr o
    at main.lhs:(824,1)-(825,21)
  Expected type: Expr i
    Actual type: Expr l
- In the third argument of 'substituteExpr', namely 'body'
  In the expression: substituteExpr scope' subst body
  In an equation for 'body'':
      body' = substituteExpr scope' subst body
\end{verbatim}

This kind of error message certainly doesn't teach the foil, but it does, once one learns to read it, indicate what the problem is.
To wit, we're trying to substitute \cd{body}, which has type \cd{Expr l}, with a substitution of type \cd{Substitution Expr i o}, and we shouldn't do that, because \cd{body} may refer to names that we did not define a substitution for.

\section{Discussion}

\paragraph{Other expression representations}  Our discussion has been confined entirely to representing expressions with binders in terms of explicit names.
Other expression representations exist---the more popular ones are De Bruijn indices \citep{de-bruijn}, implemented in Haskell by Ed Kmett's excellent Bound library \citep{bound-code, bound-blog}; the locally-nameless representation \citep{locally-nameless}; and variants of higher-order abstract syntax \citep{hoas}.

None, however, completely avoid the fundamental issue of name capture, as evidenced by the length and incompleteness of the preceding list.
We refer the curious reader to \citet{lambda-n-ways, keynote} for recent and ongoing discussions on the relative merits of expression representations.
Our contribution with the foil is to show that explicit names can be not only fast and stateless, per the rapier \citep{secrets}, but also relatively error-free.

Many similar tricks (such as tagging expressions by their scopes) can be found in compilers written in richer, dependently-typed host languages \citep{idris2, idris-implementation-docs}.
Our contribution here is to show how similar approaches can be productively embedded in a language with a more classical type system such as Haskell.

And it turns out that the way we embed those approaches resembles the ideas of \citet{ghostsofproofs}, only specialized to proving the scoping properties of terms.
We also use phantom type variables to attach type-level identifiers to expressions, in our case denoting the scopes they're in.
Our work can be seen as a productive application of that approach, beyond the ones presented in the original work.

\paragraph{Parsing and name resolution}
If the object language is not itself embedded in Haskell, how does one get an \cd{Expr n} indexed by the correct scope to start applying compiler passes to?
Using the foil implies an explicit \emph{name resolution} pass near the beginning of a compiler's pipeline, looking something like this:
\begin{code}
resolveNames :: (Distinct n) => Scope n -> Map String (Name n)
  -> UExpr -> Expr n
resolveNames scope env = \case
  UVar str -> case M.lookup str env of
    Just name -> Var name
    Nothing -> error ("Unbound variable " ++ str)
  UApp f x -> App (recur f) (recur x)
    where recur = resolveNames scope env
  ULam str body -> withFresh scope \binder ->
    let scope' = extendScope binder scope
        env' = M.insert str (nameOf binder) (fmap sink env)
        body' = resolveNames scope' env' body in
    Lam (LamExpr binder body')
\end{code}

The idea is that \cd{UExpr} is a variant of \cd{Expr} that contains strings (or whatever the previous stage of parsing produces) instead of foil-managed \cd{Name}s, and we carefully implement one substitution-like function to convert them.
Note that this function can fail if the input program was not scope-correct; but if it succeeds, the foil will guarantee scope correctness of all downstream compiler passes.

The \cd{resolveName} function itself is only partly protected by the foil: forgetting to extend the scope (and thus producing spurious shadowing) doesn't type-check in Haskell, but forgetting to extend the \cd{env} map (and thus producing a spurious unbound variable error) does.
However, this is just one function, and dealing with names is its whole focus, so the error surface is much smaller than it would be without the foil.

\paragraph{How shadowing happens}
All the names the foil generates are fresh relative to the enclosing scope.
However, they need not be globally distinct, so sinking one binding form past another may introduce shadowing if the bound variable happens to be the same.

\paragraph{Name uniqueness invariants}  The foil allows name shadowing in expressions at rest---there is nothing stopping a \cd{NameBinder n l} that appears as the binder in a \cd{LamExpr n} from shadowing one of the names in the \cd{n} scope.
We pay for this during substitutions, by having to (i) check whether such shadowing actually happens, and (ii) renaming that name if so.

Perhaps we could save ourselves the trouble by forbidding shadowing at rest, i.e., requiring that all scopes are always \cd{Distinct}?
That would save us from renaming a binder that clashes with the free variables of the substituend, because that would never happen.
However, we would still have to contend with name clashes, because to maintain the no-shadowing invariant we would have to check whether a binder collided with any \emph{bound} variables of the term being sunk.
Perhaps a variant of the foil could be developed to make that safe, but it's not clearly advantageous.

Can we go further in this vein and just require all names to be globally unique?
That would certainly prevent name clashes, but still comes at a cost: now, duplicating an expression (e.g., to inline it to more than one site) would require work renaming all the binders in the copy to keep them distinct from the binders in the original.
Making \emph{that} discipline safe seems to require the expression to be typed linearly in the host language, and again for an unclear performance profile relative to the shadow-permitting rapier.

\paragraph{Distinctness constraints}  The \cd{Distinct} constraint we introduced in \Cref{sec:safer-sinking} is arguably too strong: the only thing we need to be able to sink is knowing that the new names introduced in the extension from \cd{n} to \cd{l} do not shadow names in \cd{n}.
The assertion that all the names in \cd{l} are distinct certainly implies this; would the foil work with a more precise version of the Distinctness Invariant?

One attempt would be to use \cd{Distinct n l} to mean that no names introduced between \cd{n} and \cd{l} shadow each other.
This representation, however, is not transitive! It can be easily seen that \cd{Distinct a b} and \cd{Distinct b c} doesn't imply \cd{Distinct a c}: consider two adjacent binders with the same name.

One way to fix the transitivity problem is to additionally require \cd{Distinct n l} to prove that no names introduced between \cd{n} and \cd{l} shadow \cd{Scope n} (but say nothing about distinctness in \cd{Scope n} itself).
This would be less stringent than the invariant used in our presentation, and would be sufficient to prove \cd{sink} safe.
However, the extra condition still requires reasoning about the entire scope and hence does not seem to simplify the implementation significantly (while introducing even more type parameters to think about).

\paragraph{Abstract binder types}  Similarly to how we have generalized \cd{sink} to work over arbitrarily (\cd{S -> *})-kinded types, in practice it is often useful to use a richer set of binders than only \cd{NameBinder}. For example, binder pairs:
\begin{code}
data PairB (b1 :: S -> S -> *) (b2 :: S -> S -> *) (n::S) (l::S)
  where PairB :: b1 n h -> b2 h l -> PairB b1 b2 n l
\end{code}
Just like all expression-like things have an \cd{S -> *} kind, all binder-like things have an \cd{S -> S -> *} kind (e.g. \cd{PairB} does not act like a binder, but \cd{PairB NameBinder NameBinder} does).

\paragraph{Generic implementations}  We have found that in practice it is possible to define a small-ish language of expression and binder combinators (\cd{PairB} is one example), a combination of which can express the naming discipline used in other custom types used in language syntax trees.
This trick has an added benefit that many of the typeclasses used by the foil (substitutability, sinkability, \ldots) are derivable generically once an isomorphism from a custom type to a composition of those combinators is specified.

\paragraph{Other useful operations}  In this work we have focused on substitution (and sinking as its crucial component), but those are not the only useful operations to perform on expressions and binders.
The foil can of course be used to express those as well.
To give two examples, we display the types of two such functions. Here is a type for a function that hoists an expression above a binder (but might fail if the expression mentions the bound name):
\begin{notcode}
hoist :: _ => b n l -> e l -> Maybe (e n)
\end{notcode}
This function exchanges two binders (but again might fail, if the lower binder has the other name free e.g. in its annotation):
\begin{notcode}
exchangeBinders :: _ => PairB b1 b2 n l -> Maybe (PairB b2 b1 n l)
\end{notcode}
In both cases we omit typeclass constraints for simplicity.

Without the discipline of the foil, \cd{hoist} in particular looks like the identity function, and is very easy to forget.
For example, consider inferring the type of a lambda expression.
In a dependently typed object language, the inferred type of the result could, in principle, depend on an intermediate value; but the type of the whole lambda expression must only depend on the argument, and on variables in scope where the lambda is defined.
Type inference must therefore check for such leaks and deal with them.
The \cd{hoist} function performs this leak check, by testing whether any of the names free in the \cd{e l} argument are bound by the \cd{b n l} argument; and the foil reminds the user that they need to apply it to reconcile scope indices.

\paragraph{Builder monads}  While in all examples provided here, we've manually used the low-level naming implementation, elaborations in large languages (such as Dex) are very conveniently expressed in a monadic style, where the monad is responsible for building up a program based on the ``emitted instructions''.
For example:
\begin{notcode}
emit :: Expr -> Builder Var

-- Simplify the expression to a fully
-- evaluated value, or emit a simpler expr.
simplify :: Expr -> Builder (Either Var Value)
simplify expr = case expr of
  Multiply x y -> case (x, y) of
    (Lit xl, Lit yl) -> return $ Right $ xl * yl
    (_     , Lit 2 ) -> liftM Left $ emit $ ShiftLeft x (Lit 1)
    _                -> liftM Left $ emit $ Multiply x y
  ...
\end{notcode}
However, having developed the presented naming system, we would like the terms built by the monad to always be well-scoped.

While not trivial, this (unsurprisingly?) can be achieved through the use of \cd{S}-indexed monads.
Instead of \cd{Builder a}, we would use \cd{Builder n a}, which would place an additional namespace restriction on \cd{emit}:
\begin{notcode}
emit :: Expr n -> BuilderM n (Var n)
\end{notcode}

An interesting concern when using such a monad is that the type parameter \cd{n} is in fact \emph{mutable}: \cd{emit} modifies the scope by binding the expression to a fresh variable, but it still runs in \cd{n}!
What happens here is that all \cd{n}-scoped values are \emph{implicitly sunk}, which is why it is so important for \cd{sink} to have no effect on the run-time representation of terms.

Of course, one has to be careful so that no non-sinkable \cd{n}-indexed values can be provided to the user, but since the names are generated by the monad, this can be done by restricting its interface.
For example, it shouldn't be possible to ask \cd{Builder n a} for \cd{Scope n}, but it is ok to provide a \cd{Builder} method for querying whether a given name is in scope: its result type, \cd{Bool}, is (trivially) considered sinkable.

\section{Conclusion}

We presented the foil, a technique for managing explicit names in a program representation that is fast, stateless, and hard to misuse.
Speed and statelessness come from the foil being an exact reimplementation of the rapier from \citet{secrets}; our addition was to spell out what invariants correct use of the rapier requires, and to use a phantom type to get a Haskell type-checker to enforce them.
Because the type is phantom, adjusting types where it is safe is done by \cd{unsafeCoerce}, so imposes no runtime cost.
We hope that future (or current) compilers written in Haskell (or another language with a sufficiently powerful type system) can use the foil to avoid name-handling bugs.

\bibliographystyle{ACM-Reference-Format}
\bibliography{main}

\appendix

\section{Expression Sinkability}
\label{sec:expr-sinkability}

\begin{code}
extendRenaming :: (Name n -> Name n') -> NameBinder n l
  -> (forall l'. (Name l -> Name l') -> NameBinder n' l' -> r)
  -> r
extendRenaming _ (UnsafeBinder name) cont =
  cont unsafeCoerce (UnsafeBinder name)
\end{code}

\begin{figure}
\begin{code}
instance Sinkable Expr where
  sinkabilityProof rename (Var v) = Var (rename v)
  sinkabilityProof rename (App f e) =
    App (sinkabilityProof rename f) (sinkabilityProof rename e)
  sinkabilityProof rename (Lam lam) =
    Lam (sinkabilityProof rename lam)

instance Sinkable LamExpr where
  sinkabilityProof rename (LamExpr binder body) =
    extendRenaming rename binder \rename' binder' ->
      LamExpr binder' (sinkabilityProof rename' body)
\end{code}
\caption{Sinkability proof for \cd{Expr}}{This is boilerplate, and could be generated based on a variant of the \cd{Generic} class for (\cd{S -> *})-kinded types.}
\label{fig:expr-sinkability}
\end{figure}

The sinkability proof for our lambda calculus \cd{Expr n} is in \Cref{fig:expr-sinkability}.
The only subtlety is going under a binder in a sinking proof, which requires extending the renaming map to apply to the local scope.
Luckily that subtlety is generic, so the \cd{extendRenaming} function below need only be implemented once.

We know of no safe implementation for this function, but here is the argument for why the function (with the given type) is safe even if implemented unsafely.

Let the name in the binder be \cd{x}.
The scopes are related thus:
\begin{notcode}
l = n ++ [x]
l' = n' ++ [x]
\end{notcode}
We are given a renaming from \cd{n} to \cd{n'}, and wish to produce a renaming from \cd{l} to \cd{l'}.
Any name in \cd{l} must be either
\begin{enumerate}
    \item \cd{x} itself, in which case it's also in \cd{l'}, or
    \item in \cd{n}, in which case it can be renamed to \cd{n'}.  The only issue would be
       if it were shadowed by \cd{x}, but it can't be because then we'd be in case (1).
\end{enumerate}
The resulting renaming itself is of course irrelevant, because the only purpose of sinkability proofs is to be type-checked.

\end{document}